\documentclass[peerreview, draftcls, onecolumn, 12pt]{IEEEtran}
\usepackage{graphics}
\usepackage{amsmath, amsthm, amssymb}
\usepackage[dvips]{graphicx}
\usepackage{ifpdf}
\usepackage{cite}
\usepackage{xspace}
\usepackage{color}

\newcommand{\EXPC}[3]{\mathbb{E}_{#1}\left(\left.{#2}\right|{#3}\right)}

\newcommand{\PROB}[1]{\mathcal{P}\left(#1\right)}
\newcommand{\eqn}[2]{\begin{equation}\label{#1}#2\end{equation}}

\begin {document}
\title{Separating the Effect of Independent Interference Sources with Rayleigh Faded Signal Link: Outage Analysis and Applications}
\author{
Arshdeep~S.~Kahlon,
Sebastian~S.~Szyszkowicz,~\IEEEmembership{Member,~IEEE,}
Shalini~Periyalwar, ~\IEEEmembership{Senior Member,~IEEE,}
Halim~Yanikomeroglu,~\IEEEmembership{Member,~IEEE}
\thanks{Manuscript submitted January 24, 2012. The authors are with the Department of Systems and Computer Engineering, 
Carleton University, Ottawa, Ontario, Canada.
E-mail: \{akahlon, sz, shalinip, halim\}@sce.carleton.ca}
}
\maketitle
\begin {abstract}
We show that, for independent interfering sources and a signal link with exponentially distributed received power, the total probability of outage can be decomposed as a simple expression of the outages from the individual interfering sources. 
We give a mathematical proof of this result, and discuss some immediate implications, showing how it results in important simplifications to statistical outage analysis. We also discuss its application to two active topics of study:  spectrum sharing, and sum of interference powers (e.g., lognormal) analysis.
\end {abstract}
\begin {IEEEkeywords}
wireless interference, outage probability, Rayleigh fading, spectrum sharing, heterogeneous networks
\end {IEEEkeywords}

\section{Introduction}

With the increasing need for spatial spectrum reuse and for co-channel coexistence of heterogeneous wireless networks, the effect of the combined interference from multiple sources is becoming an important topic of study. While this problem has received several decades of theoretical study under various research directions, it remains analytically challenging, largely due to the need of finding the distribution of the sum of the random interference powers \cite{Win_Pinto_Survey}, notably when they are lognormally distributed\cite{AA0294, NB0304, MD0909, CT0510, SS9}. 

There may be cases when we wish to study the outage at a receiver due to the sum of independent interfering signal powers, yet the distribution of the constituent interfering powers is unknown. Such a case can be considered in a spectrum sharing scenario
where two or more heterogeneous networks share the same spectrum \cite{DSAsurvey,JsacCellularAdhocbounds,Thompson_aggregateinterference,PIMRC1ASSH}.
Throughout this paper, we only consider spectrum sharing without any spectrum sensing or cognition, which implies that the secondary network
necessarily increases the outage probability of primary receivers.
The operator of the primary network may be interested in obtaining insights into the additional outage that a receiver would suffer from the deployment of a heterogeneous secondary network, in order to determine the feasibility of spectrum sharing. 

In this work, we show that in the case of independent interfering powers following any distribution, and an independent signal power with exponential power distribution, it is possible to separate the outage effect of each interferer. We show this result analytically and exactly, and discuss some of its more immediate consequences for the simplification of outage analysis.

In Section II, we give the general outage problem as it is often formulated. In Section III, we introduce our main
expression for the total outage probability and the mathematical result it is based on, and make some general observations on its consequences to outage analysis. In Section IV, we show how our result can concretely simplify calculations in two important research topics: 1) primary/secondary network sharing scenarios and 2) sum of lognormals modeling, before concluding in section V.

\section{Problem Formulation}
Consider a wireless device receiving a useful signal with power $S$, and suffering from a total received interference of power $I$. We assume that $S$ is exponentially distributed (due, notably, to Rayleigh fading), while $I$ can be written as
\eqn{eqn_I}{
I = \sum_{i=1}^N I_i,
}
where $\{I_i\}$ is the set of the $N$ independent received interference powers (which may originate from individual transmitters, entire networks or parts thereof, or thermal noise), and the interference powers are assumed to add incoherently (in power) \cite{Win_Pinto_Survey, AA0294, NB0304,  CF0308, SS9}, and are treated as additive white Gaussian noise as far as outage is concerned \cite{AA0294, CF0308}. We also assume the signal power to be independent of the interference powers.

The outage probability $\varepsilon$ on the signal link is obtained from the cumulative distribution function (CDF) of the power ratio $S/I$:
\eqn{eqn_eps}{
\varepsilon=\PROB{\frac{S}{I}<\beta}=\PROB{\frac{S}{\sum_{i=1}^N I_i}<\beta},
} 
where $\beta$ is the outage threshold in terms of the signal to interference (and noise) power ratio.

\section{Analysis}
\subsection{Mathematical Result}
We introduce a result on random variables (RVs) that will allow us to separate the summation of interference powers under our assumptions. Consider $\{X_i\}$ a set of $N$ independent RVs, and $Y$ an independent exponentially distributed RV. We may then write
\eqn{eqn_XiY}{
\PROB{\sum_{i=1}^NX_i<Y}=\prod_{i=1}^N\PROB{X_i<Y}.
}
The proof is in the Appendix.

\subsection{Separability of the Interference Powers}
Applying (\ref{eqn_XiY}) to the outage expression (\ref{eqn_eps}), identifying $X_i=I_i$ and $Y=S/\beta$, and inverting the inequalities gives
\eqn{eqn_separableI}{
\varepsilon=1-\prod_{i=1}^N\left(1-\varepsilon_i\right),\quad
\varepsilon_i=\PROB{\frac{S}{I_i}<\beta}.
}
We have thus expressed the total outage probability $\varepsilon$ as a simple algebraic expression of the partial outage probabilities $\varepsilon_i$ that would have been caused by each individual interfering source separately (given the same outage threshold $\beta$).
This generalises (and simplifies) similar results in \cite{Yao_Sheikh_TVT1992} and in \cite{VG0811} (and references therein), where \textit{all} the signals are Rayleigh faded.

\subsection{Some Useful Consequences}
Some interesting observations immediately result from (\ref{eqn_separableI}):
\begin{enumerate}
\item The difficulty of finding the CDF of the ratio of the sum of $N$ independent RVs and the exponentially distributed RV disappears, and reduces to that of finding $M$ CDFs of the ratio of that exponentially distributed RV and each independent RV, where $M\leq N$ is the number of statistically different RVs (if all the sources have the same statistics, then $M=1$).
\item The total outage probability can be obtained directly from the partial outage probabilities, without the need to know the models of the underlying interferences, which is useful when the partial outage probabilities are obtained from simulation, field measurements, or even usage statistics of a working network. 
\item If the interference powers are statistically dependent (due, e.g., to correlated shadowing \cite{AA0294, MD0909, SS9}), our result can still be useful if the interferers can be grouped in such a way that the interferences are independent across the groups. Then, in order to calculate the outage probability, the CDFs of the sum interferences (or the partial outage probabilities) need to be found only within those groups, but not globally.
\end{enumerate}

Our result has important implications in simplifying the analysis of outage caused by multiple interference sources. In the next section, we show how our result can concretely be applied to two research directions that have already received much attention.

\section{Applications to Current Research}
Our result in (\ref{eqn_separableI}) has immediate applications in simplifying various outage calculations, e.g., when a secondary network shares the spectrum without sensing the primary network's activity, thereby increasing the outage probability. The result also simplifies outage probability calculations when the interference is modeled as the sum of independent lognormal RVs. 

\subsection{Spectrum Sharing between Primary and Secondary Network}
A direct application of our result with $N=2$ can be seen in the spectrum sharing
scenario, where we want to find the additional outage at a primary network receiver due to the deployment of
a secondary network, while avoiding the potentially complex task of characterising the interference
from the primary network. 

Consider a typical primary receiver, experiencing an outage probability $\varepsilon_1$ due to co-channel interfering primary transmitters (in the absence of the secondary network). We call $\varepsilon_{\mathrm T}$ the maximum outage probability allowed at a primary receiver. It then follows from (\ref{eqn_separableI}) that the secondary network must be designed in such a way that the outage probability $\varepsilon_2$ caused by its interference alone (in the absence of the primary network's co-channel interferers) satisfies
\begin {equation}
\varepsilon_2\leq\frac{\varepsilon_{\mathrm T}-\varepsilon_1}{1-\varepsilon_1}.
\end {equation}  

\subsection{Outage Analysis Using the Sum of Lognormal Random Variables}
The study of the distribution of the sum of several lognormal RVs has received much research attention for several decades \cite{AA0294, NB0304}, and still attracts significant interest \cite{MD0909, CT0510, SS9}. The research is primarily (but not exclusively) in the field of wireless communications, where it is motivated by the model where each interference source suffers (possibly correlated) lognormal shadowing. In this case, each $I_i$ is modeled as a lognormal RV, and the challenge is to find the sum distribution of $I$. However, no closed-form solution exists \cite{MD0909, CT0510} even for the simplest cases, and in fact there exist many approximating methods that trade-off accuracy against simplicity.

It is important to see the context of this research: the goal of finding the sum distribution of the interference is not necessarily an end in itself. Its main use is as an intermediate step in finding the distribution of the signal-to-interference-(and possibly noise)-power ratio, and hence the outage probability \cite{AA0294, NB0304, CF0308}. Our result (\ref{eqn_separableI}) shows that, given independent interference powers and an exponentially distributed received signal power, the unsolved problem of a sum of lognormal RVs disappears, and essentially reduces to the problem of the outage from a single lognormal interferer:
\begin{equation}
\varepsilon_i=\PROB{\frac{S}{I_i}<\beta}=\PROB{\sqrt{S}\cdot{I_i}^{-1/2}<\sqrt{\beta}}.
\end{equation}
Now, $\sqrt{S}$ follows a Rayleigh distribution, while ${I_i}^{-1/2}$ is an independent lognormal RV, hence the problem reduces the computation of $M\leq N$ probabilities from the Suzuki distribution\footnote{This is a well established numerical calculation: e.g., the {SuzukiDistribution}$[\mu,\nu]$ function in Wolfram \emph{Mathematica}.}, given $M$ statistically distinct lognormal RVs.

The result can also be extended to the case where the interference powers are not lognormal: notably they may include small-scale fading (lognormal-times-fading power, as in \cite{CF0308}), and path loss based on random positions \cite{SS9}. It remains the case that the most difficult probability calculation, i.e., the summation of random powers, need not be performed.

\section{Conclusion}
We showed that, under the assumption of independent received interference powers, and an exponentially distributed received signal power (e.g., due to Rayleigh fading), the outage probability due to all the interfering sources can easily be decomposed into the partial outage probabilities as would be caused by the interferers individually. It is therefore not necessary to know the distribution of the total interference power to find the outage probability of the system, nor in fact even that of the individual interference powers, as long as the corresponding partial outage probabilities are known.

Our result makes important simplifications in the calculation of outage probability, which is applicable in a variety of
scenarios, and notably in the case of spectrum sharing, as well as in the case of sum of lognormal RVs interference modeling. It naturally extends to include noise powers as well.
It has the advantage of being simple and exact, and can be used in practical scenarios with possibly complex and intractable interfering sources, in order to get insights into the effects of those various interference sources. 

\section*{Appendix}
\begin {proof}
[Proof of (\ref{eqn_separableI})]
We can write the left hand side of \eqref{eqn_XiY} as
\eqn{Pr1}
{
\EXPC{X_1,X_2,\ldots,X_N}{\PROB{\sum_{i=1}^NX_i<Y}}{X_1,X_2,\ldots,X_N}.
}
Let $\mu$ be the mean of the exponentially distributed RV $Y$. Then we can write the above as
\eqn{Pr2}
{\begin {aligned}
&\mathbb{E}_{X_1,X_2,\ldots,X_N}\left(\exp\left( -\mu \sum_{i=1}^NX_i\right)\right)\\
&=\mathbb{E}_{X_1,X_2,\ldots,X_N}\left(\prod_{i=1}^{N}\exp{\left(-\mu X_i\right)}\right).
\end {aligned}
}
Now, since $X_1,X_2,\ldots,X_N$ are independent RVs, we can write the above as
\eqn{Pr3}
{\begin {aligned}
\prod_{i=1}^{N}\mathbb{E}_{X_i}\left({\exp{\left(-\mu X_i\right)}}\right),
\end {aligned}}
which is equivalent to
\eqn{Pr4}
{\begin{aligned}
\prod_{i=1}^{N}\PROB{X_i<Y}.
\end {aligned}
}
\end {proof}

\bibliographystyle{IEEEtran}
\bibliography{References} 

\end{document}